\documentclass[aps,prl,twocolumn,amsmath,amssymb]{revtex4}
\usepackage{dcolumn}
\usepackage{bm}
\usepackage{graphicx}
\begin{document}
\title{Magnetic glasses and structural glasses: devitrification and a reentrant transition under CHUF protocol.}
\author{P. Chaddah and A. Banerjee}
\affiliation{UGC-DAE Consortium for Scientific Research, University Campus, Khandwa Road, Indore-452001, Madhya Pradesh, India.}

\begin{abstract}
A recent paper from Raveau's group asserts that the specially designed CHUF measurement protocol serves to bring out a special feature of the magnetic glass state. This protocol, enunciated and applied in our publications since over three years, allows establishing phase coexistence through macroscopic measurements and distinguishing the metastable and stable phases (amongst the coexisting phase fractions across a first order magnetic transition) of a glass-like arrested state. In view of the recent report of the vitrification of monoatomic germanium under pressure, we discuss the applicability of an analogous CHUP protocol for states across an arrested first order structural transition, and specifically in establishing whether the vitrification was partial or complete.  
\end{abstract}

\maketitle
Phase-coexistence has been observed to persist in many half-doped manganites, across a first order magnetic transition, down to the lowest temperature. Based on studies on doped CeFe$_2$, this was conjectured as resulting from kinetic arrest analogous to that seen in a quenched metallic glass \cite{manekar}. It was pointed out that cooling in different values of magnetic field may aid or prevent this kinetic arrest; cooling in one value of magnetic field may lead to a glass-like arrested state (GLAS), while cooling in a different field may allow the first order transition to be completed and the equilibrium state to be established. Since the first order transition is broadened, occurring over regions of the length scale of the correlation length, the transition was argued \cite{ban1} to be partial for cooling field lying between two values (H$_1$ and H$_2$) of magnetic field, resulting in coexisting phases. For cooling fields below H$_1$ the transition was completed (totally arrested) if the high-temperature phase was ferromagnetic (antiferromagnetic), and for transition fields above H$_2$ the transition was totally arrested (completed) if the high-temperature phase was ferromagnetic (antiferromagnetic). It was argued \cite{ban1} that by using different cooling and warming fields (H$_C$ and H$_W$), de-arrest (or devitrification) of the kinetically arrested phase could be caused, and this de-arrest would be seen for only one sign of (H$_C$ - H$_W$). Further heating would cause this de-arrested state to undergo the reverse magnetic transition, and CHUF would show a reentrant transition. The sign of (H$_C$ - H$_W$) for which a reentrant transition is observed on warming would depend on the magnetic order in the equilibrium low-temperature phase, and this was demonstrated \cite{ban1}. The protocol `cooling and heating in unequal fields' with the acronym CHUF was specified, its potential outlined, and it was applied in studies on many magnetic materials showing kinetic arrest \cite{arxiv, ban2, pallavi, roy}. The technique of using CHUF with opposing signs of (H$_C$ - H$_W$) has immense applicability in materials that show phase-coexistence, of two competing magnetic phases, down to the lowest temperature. The CHUF protocol has been applied, beyond bulk half-doped manganites where it originated \cite{ban1, ban2}, to thin films and bulk samples of various intermetallics with functional relevance \cite{pallavi}. The concept of phase coexistence resulting from kinetic arrest has been accepted as an alternative explanation to sophisticated theories considering the phase coexistence as an equilibrium state \cite{shenoy}.  The recent work from Raveau's group \cite{sarkar} uses this well-established (and no longer `specially designed'!) protocol in a new family of magnetic glass, and also shows tunability of the phase fractions in the state of phase-coexistence \cite{ban1, arxiv2}. The power of CHUF lies in showing whether there is coexistence of an equilibrium phase and a GLAS (on cooling in H$_C$) without the need for microscopic measurements, in establishing that the state of phase-coexistence is not an equilibrium state, and in identifying which of the coexisting phases is the arrested phase \cite{ban2, pallavi}.  

It has been argued often \cite{arxiv} that applying a variable H does not require a medium, and this makes it experimentally far more tractable than applying a variable pressure (which does require a medium). Moreover, studies on glass-formation have recently benefited by the use of the `magic ingredient' of pressure \cite{tar}, when Bhat et al. \cite{bhat} could vitrify monoatomic germanium by rapid-cooling under high pressures (above 7.9 GPa). The discussion above on the CHUF protocol has exploited the use of magnetic field as the `magic ingredient'.

The CHUF protocol in magnetic glasses allows the observation of vitrification on cooling, and devitrification on heating, with both heating and cooling rates being same, but with cooling field H$_C$ and heating field H$_W$ being chosen appropriately. This contrasts with the well-known case of metallic glasses where vitrification and devitrification are observed at the same value of pressure (counterpart of field for these structural transitions), but with cooling rate (for causing vitrification) being much faster than the heating rate for observing devitrification. Specifically, devitrification at T$_X$ in a DSC scan is observed with a heating rate that is at least an order of magnitude slower than the cooling rate for glass formation \cite{ino}, and this devitrification is followed by melting of the crystalline state on further heating. Since the density of the glass is close to that of the liquid, a reentrant transition of density is observed on warming at slower rate.  We propose that the `magic ingredient' of pressure could be used to observe vitrification and devitrification at the same (or similar) rates of temperature variation. Cooling under high pressure aids glass-formation (or arrest) if the transition temperature falls with increasing pressure \cite{bhat}. This happens when the high-temperature phase is of higher density, and this is similar to the case where the high-temperature phase was ferromagnetic. The critical cooling rate for vitrification reduces. The reduction in transition temperature for structural transitions is not as drastic as seen in ferromagnetic-to-antiferromagnetic transitions with achievable magnetic fields, but the possibility exists that cooling and heating in unequal pressures (CHUP) protocol could allow both vitrification and devitrification dynamics to be studied. For materials that are denser on solidification, critical cooling rate for vitrification would be faster at higher pressure and CHUP protocol would require cooling at low pressure and warming in higher pressure to observe a reentrant transition. One relevance of CHUP to the pioneering work on germanium \cite{bhat} is that it could establish whether or not there is phase coexistence, i.e. whether the vitrification under the pressure used is partial (with some fraction of crystalline germanium), or is complete.

CHUP technique can also be used to create different fraction of arrested and equilibrium phases as has been done in manganites using the CHUF protocol \cite{ban3}. This glass-ceramic state has applications because of better mechanical properties \cite{greer} and the dimension of the crystallites may be controlled as their nucleation temperatures would be different \cite{arxiv3}.

\end{document}